\documentclass[a4paper,11pt]{article}
\pdfoutput=1 

\usepackage{jinstpub} 
\usepackage{graphicx}
\usepackage{caption}
\usepackage{subcaption}
\usepackage{amsmath}   
\usepackage{xspace}

\usepackage[printwatermark]{xwatermark}
\usepackage{xcolor}
\usepackage{lipsum}
\usepackage{tikz}
\usepackage{lineno}
\usepackage{lipsum}
\usepackage{soul}

\title{Time performance of a triple-GEM detector at high rate}
\author[a,b]{A. Amoroso}
\author[c]{, R. Baldini Ferroli}
\author[d,e]{, I. Balossino}
\author[c]{, M. Bertani}
\author[d]{, D. Bettoni}
\author[a,b]{, A. Bortone}
\author[c]{, A. Calcaterra}
\author[c]{, S. Cerioni}
\author[a]{, W. Cheng}
\author[d]{, G. Cibinetto}
\author[d]{, A. Cotta Ramusino}
\author[a]{, F. Cossio}
\author[a]{, M. Da Rocha Rolo}
\author[a,b]{, F. De Mori}
\author[f]{, A. Denig}
\author[a,b]{, M. Destefanis}
\author[e]{, J. Dong}
\author[d,g]{, F. Evangelisti}
\author[1,d]{, R. Farinelli \note{Corresponding author.}}
\author[a,b]{, L.~Fava}
\author[c]{, G. Felici}
\author[f]{, B. Garillon}
\author[d,g]{, I. Garzia}
\author[c]{, M. Gatta}
\author[a]{, G. Giraudo}
\author[d,g]{, S. Gramigna}
\author[a,b]{, M. Greco}
\author[f]{, P. Gülker}
\author[f]{, Y. P. Guo}
\author[f]{, W. Lauth}
\author[a,b]{, L.~Lavezzi}
\author[a,b]{, M. Maggiora}
\author[d]{, R. Malaguti}
\author[h,i]{, A. Mangoni}
\author[a,b]{, S. Marcello}
\author[d]{, M.~Melchiorri}
\author[d,e]{, G. Mezzadri}
\author[c]{, E, Pace}
\author[h,i]{, S. Pacetti}
\author[c]{, P. Patteri}
\author[2,a,b]{, J. Pellegrino \note{Currently at: right. based on science GmbH.}}
\author[f]{, C. F. Redmer}
\author[f]{, M. Ripka}
\author[a]{, A.Rivetti}
\author[f]{, C. Rosner}
\author[d,g]{, M.~Scodeggio}
\author[a,b]{, S. Sosio}
\author[a,b]{, S. Spataro}

\affiliation[a]{INFN, Sezione di Torino, via P. Giuria 1, 10125 Torino, Italy}
\affiliation[b]{Universit\`a di Torino, Dipartimento di Fisica, via P. Giuria 1, 10125 Torino, Italy}
\affiliation[c]{INFN, Laboratori Nazionali di Frascati, via E. Fermi 40, 00044 Frascati (Roma), Italy}
\affiliation[d]{INFN, Sezione di Ferrara, via G. Saragat 1, 44122 Ferrara, Italy}
\affiliation[e]{Institute of High Energy Physics, Chinese Academy of Sciences, 19B YuquanLu, Beijing, 100049, People's Republic of China}
\affiliation[f]{Johannes Gutenberg University of Mainz, Johann-Joachim-Becher-Weg 45, D-55099 Mainz, Germany}
\affiliation[g]{Universit\`a di Ferrara, Dipartimento di Fisica e Scienze della Terra, via G. Saragat 1, 44122 Ferrara, Italy}
\affiliation[h]{INFN, Sezione di Perugia, via A. Pascoli, 06123 Perugia, Italy}
\affiliation[i]{Universit\`a di Perugia, Dipartimento di Fisica e Geologia, via A. Pascoli, 06123 Perugia, Italy}
\emailAdd{rfarinelli@fe.infn.it}

\abstract{Gaseous detectors are used in high energy physics as trackers or, more generally, as devices for the measurement of the particle position. For this reason, they must provide high spatial resolution and they have to be able to operate in regions of intense radiation, i.e. around the interaction point of collider machines.
Among these, Micro Pattern Gaseous Detectors (MPGD) are the latest frontier and allow to overcome many limitations of the pre-existing detectors, such as the radiation tolerance and the rate capability. The gas Electron Multiplier (GEM) is a MPGD that exploits an intense electric field in a reduced amplification region in order to prevent discharges. Several amplification stages, like in a triple-GEM, allow to increase the detector gain and to reduce the discharge probability. Reconstruction techniques such as charge centroid (CC) and micro-Time Projection Chamber ($\upmu$TPC) are used to perform the position measurement. From literature triple-GEMs show a stable behaviour up to $10^8\,$Hz/cm$^2$. A testbeam with four planar triple-GEMs has been performed at the Mainz Microtron (MAMI) facility and their performance was evaluated in different beam conditions. In this article a focus on the time performance for the $\upmu$TPC clusterization is given and a new measurement of the triple-GEM limits at high rate will be presented.}

\arxivnumber{XXX}

\begin{document}
\maketitle
\flushbottom
\section{Introduction}

High energy physics experiments are continuously searching for new results by improving the precision of measurements and by increasing the luminosity of colliders. Future detectors need to follow step by step the evolution of the physicists' needs. 

Thanks to the technological support of the photo-lithographic manufactures of the readout and the polymide deposition on thin layer, the new kind of detectors, named Micro Pattern Gaseous Detector (MPGD), can achieve a high granularity. This reduces the dead time of the detector and improved the rate capability above 10$^8\,$Hz/cm$^2$ \cite{rate}, two orders of magnitude higher than the wire chambers. The technology under test, the triple-GEM (Gas Electron Multiplier), is mainly used to measure particle positions in high energy physics. These detectors are used as tracking devices due to the high tolerance to intense particle flux and for their convenience to be installed over large areas, $e.g$ in muon detection chambers. In this article, the performance of a triple-GEM exposed to an intense particle flux will be described.

\vspace{0.5cm}
\section{Detection and reconstruction techniques}

\begin{figure}[tbp]
    \centering
        \includegraphics[width=0.6\textwidth]{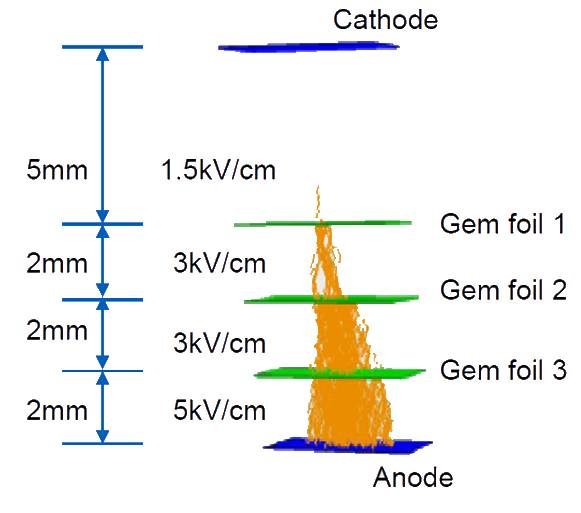}
        \caption{Representation of a triple GEM detector. Blue foils are the cathode and the anode and the green ones are the three amplification stages. On the left, the applied electric field and the distance between each electrodes are shown. The orange lines represent the path of the electrons with the opening of the avalanche inside the detector.}
        \label{fig:triple}
\end{figure}

A GEM detector amplifies the ionization generated by charged particles interacting with the gas medium that fills its volume. The multiplication stage consists of a kapton foil with a thickness of 50 $\upmu$m and copper coated faces. It has a pattern of holes with 50$\,\upmu$m diameter and 140$\,\upmu$m pitch. A voltage difference of hundreds of Volts is applied on the two copper faces and an electric field of about 10$^5\,$V/cm is generated inside the GEM holes. Electrons entering the holes gain enough kinetic energy to further ionize the gas amplifying the signal of the primary ionization. The amplification occurs only through the holes. Several stages of GEM foils allow to increase the detector gain above 10$^4$ with a small discharge probability: less than 10$^{-6}$ with $\alpha$ particles \cite{discharge}.
The full design of a standard triple-GEM detector is shown in figure \ref{fig:triple}. It consists of a cathode, three GEM foils and an anode. The GEM foils amplify the primary ionization electrons created by charged particle interacting with the gas. The electric field between the foils drives the electrons from the cathode to the first GEM, then to the others up to the anode. The electric field between the GEM faces, i.e. in the holes, is responsible for the multiplication of the electrons. The electrons generated between the cathode and the first GEM section, the drift gap, are multiplied in three stages. They are the largest contribution to the signal, about 98$\%$ \cite{poli_th}. The drift time from the first GEM to the anode is almost the same for each electron. The main differences in the time distribution are given by the origin of the primary electron in the drift gap.

The signal generated on the anode is read out by the front-end electronics that can measure the current induced by the electron avalanche, and it extracts the hit charge and arrival time. The information, together with the strip position can be used to reconstruct the incident position of the ionizing particle by means of two algorithms: the charge centroid (CC) and the micro-Time Projection Chamber ($\upmu$TPC). The algorithms are applied to a cluster of contiguous strips. The CC method averages the strip positions, weighted by their charge, while the $\upmu$TPC exploits the time information to transform the few millimeters of the drift gap in a TPC segmented in strips or pads for the signal readout \cite{atlas}. The arrival time of the signal multiplied by the drift velocity in the drift gap measures the distance of the particle path from the anode. A bi-dimensional point ($x_{hit},z_{hit}$) is associated to each firing strip, and the sequence of several points from the same cluster is fitted with a straight line to extract the particle path in the gas, hence its position. The cluster position $x_{CC}$, reconstructed with the CC, and $x_{\upmu TPC}$, reconstructed with the $\upmu$TPC, as a function of the hit position $x_{hit}$ and time $t_{hit}$, are calculated as:
\begin{equation}		
\begin{tabular}{llcrr}
 $x_{\mathrm{CC}}=\dfrac{\sum_{i}^{N_{\mathrm{hit}}}Q_{\mathrm{hit},i} \, x_{\mathrm{hit},i}}{\sum_{i}^{N_{\mathrm{hit}}} Q_{\mathrm{hit},i}}$ ,
 &  \,  &
 $x_{\mathrm{\upmu TPC}}=\dfrac{gap/2 - b}{a}$ ,
 &  \,  &
 $z_{\mathrm{hit}}=t_{\mathrm{hit}} \, v_{\mathrm{drift}}$ ,
\end{tabular}
\label{eq:CC_TPC}
\end{equation}
where $N_{\mathrm{hit}}$ is the number of hits in the cluster, called {\it cluster size}, $x_{\mathrm{hit},i}$ and $Q_{\mathrm{hit},i}$ are the hit position and charge; $gap$ is the drift gap width, $a$ and $b$ are the linear fit parameters; $v_{\mathrm{drift}}$ is the electron drift velocity in the drift gap determined from simulations.

\begin{figure}[tbp]
    \centering
        \includegraphics[width=0.6\textwidth]{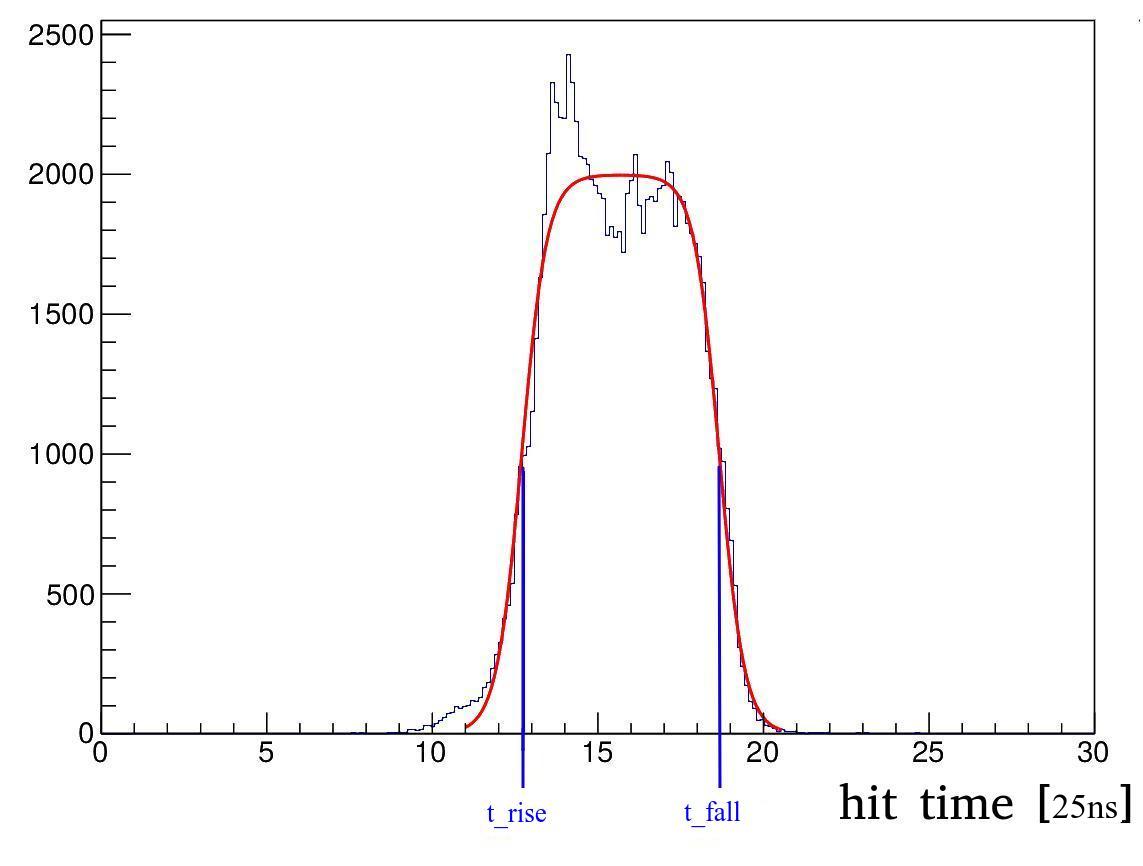}
        \caption{Time distribution of the hits in a triple-GEM with a 5 mm drift gap, Ar+iC$_4$H$_{10}$ gas mixture and APV-25 electronics, for a large number of events. Two Fermi-Dirac fits are used to describe the edges of the distribution in order to measure its width.}
        \label{fig:time_dist}
\end{figure}

\label{sec:time}
From the time distribution of all the channels, the time reference of the electron in the drift gap can be extracted. The time reference is the average time needed by an electron to drift from the first GEM to the anode, and it corresponds to the middle value of the rising edge in the time distribution. The falling edge of the time distribution corresponds to the $slowest$ primary electrons, generated closer to the cathode. The difference between the falling and the rising edge corresponds to the drift time needed by an electron to move from the cathode to the first GEM. Since the gap thickness is known, it is possible to measure the drift velocity of the electron in the drift gap as:

\begin{equation}		
v^{measured}_{drift} = \frac{gap}{t_{fall} - t_{rise}}
\label{eq:drift_time}
\end{equation}

where $t_{rise}$ and $t_{fall}$ are the time values from the rising and falling edges of the time distribution fitted by two Fermi-Dirac functions, as shown in figure \ref{fig:time_dist}.

Once the detector calibration parameter values are set, the spatial resolution achievable with a specific algorithm depends only on the angle between the trajectory of the ionizing particle and the detector; if the trajectory is orthogonal then CC provides the best measurement with a resolution of about 50$\,\upmu$m, while the $\upmu$TPC is not efficient. As the angle increases, the $\upmu$TPC reaches a resolution of 100-150$\,\upmu$m, while the performance of the CC degrades. The two algorithms use independent quantities, and a merge of the two is feasible to provide, stable performance \cite{bes}. In order to reach a good spatial resolution in a wide angular range, both algorithms are needed.

\vspace{0.5cm}
\section{Measurements a in high rate environment}
The performance in high rate environment is studied by experimental measurements and simulations \cite{charging_up}. Once the rate increases, the time between two ionizing events is compatible with the time needed by ions to evacuate the GEM hole: while electrons can leave the hole in less than a few nanoseconds, the ions take hundreds of  nanoseconds. In a high rate environment, the accumulation of ions around the GEM hole distorts the electric field and it charges up the kapton foil between the two copper faces \cite{charging_up}. This affects firstly the gain and the drift properties of the electrons, then the spatial and time resolution of the triple-GEM detector.

\begin{figure}[tbp]
    \centering
        \begin{tabular}{cc}
        \includegraphics[width=0.4\textwidth]{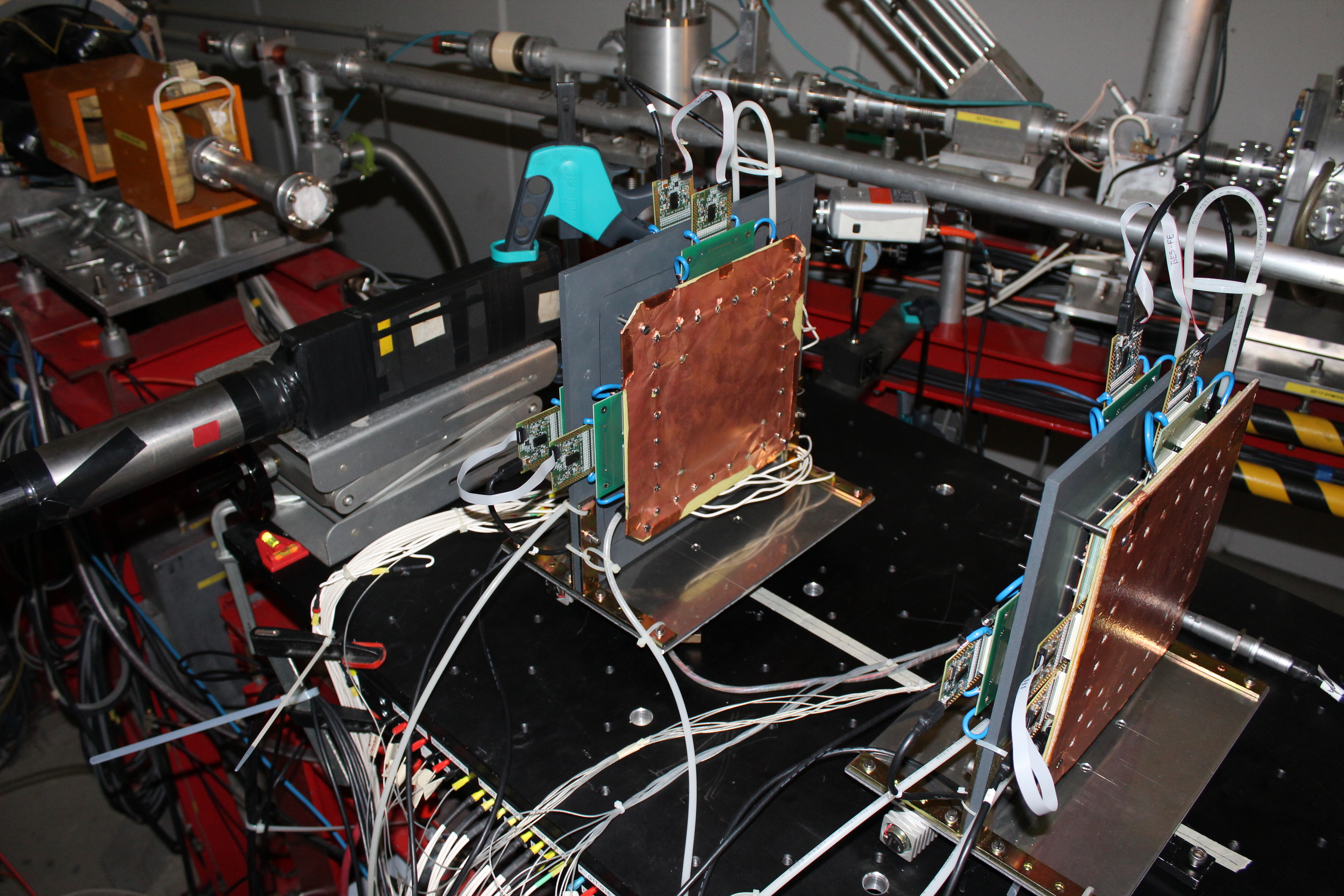}
        \includegraphics[width=0.55\textwidth]{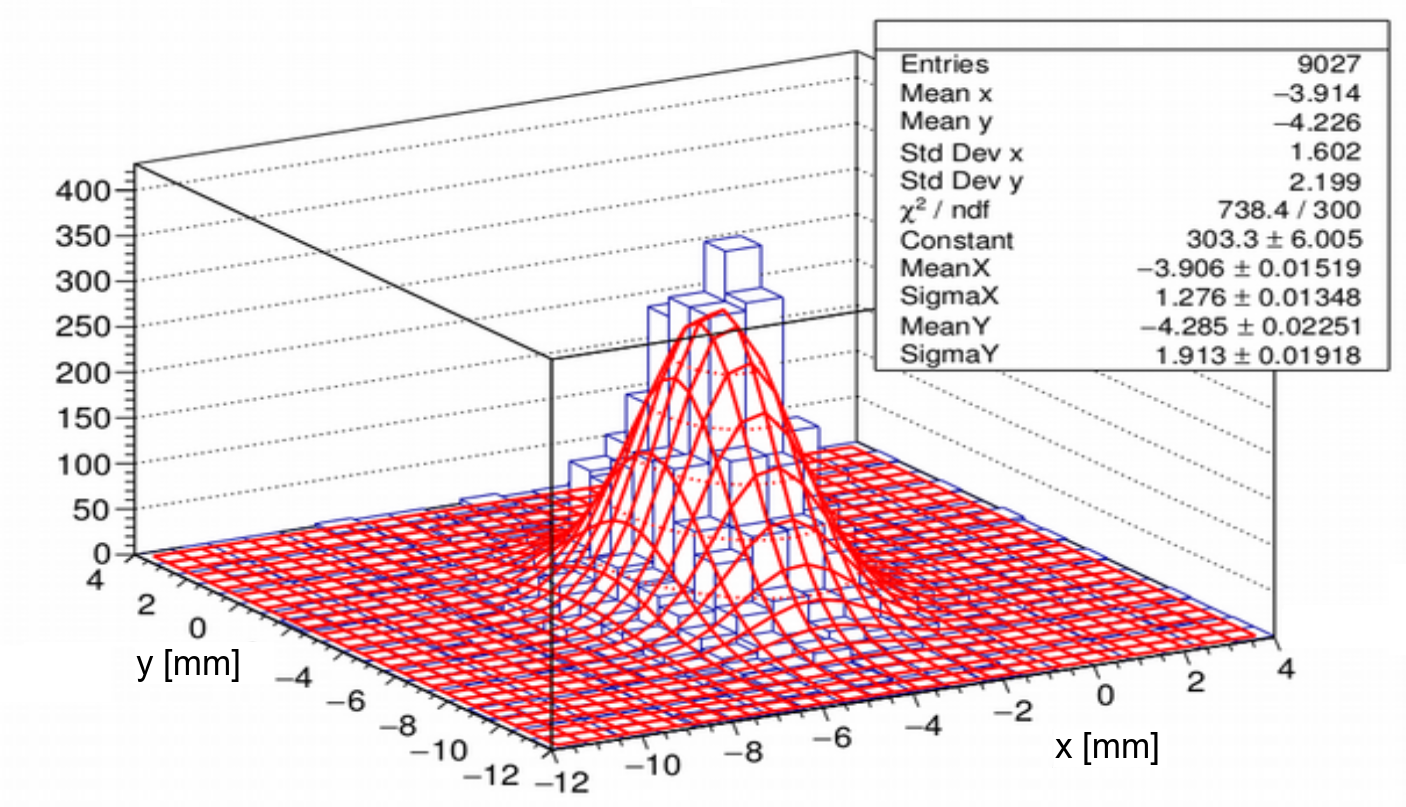}
        \end{tabular}
        \caption{On the left, a picture of the testbeam setup: from the left there are the beam line, the scintillator and the triple-GEMs. On the right, the bi-dimensional beam profile reconstructed by a triple-GEM.}
        \label{fig:setup}
\end{figure}

A testbeam at the Mainz Microtron (MAMI) facility has been performed with four planar triple-GEM detectors. The beam is composed of continuous spill of electrons emitted by a cascade of racetrack microtron and a linac injector \cite{MAMI}. The electron energy ranges from 195 MeV to 855 MeV in a spot smaller than one mm$^2$. The particle rate ranges from few kHz up to 1$\,$GHz. Two bars of plastic scintillator are used to trigger and to measure the particle flux during the experiment. The triple-GEM detectors have an active area of 10$\times$10$\,$cm$^2$ and a drift gap thickness of 5$\,$mm. The detectors are rotated by 30$^\circ$ with respect to the beam direction. The detectors are flushed with the Argon-based gas mixtures, Ar+10\%iC$_4$H$_{10}$ at first and then Ar+30\%CO$_2$. The readout planes are segmented with strips of 650$\,\upmu$m pitch on both views (XY) and are instrumented with the APV-25 \cite{apv} which can provide the charge and time information of the strip signal. The detector operates at a gain of 8000. A picture of the setup is shown in figure \ref{fig:setup} left.

The system can measure the beam rate linearly up to tens of MHz.
The triple-GEM collects on each strip of the anode a fraction of the electron avalanche, and its proper arrival time. Combining contiguous strip information it is possible to reconstruct the particle position and the beam profile, as shown in figure \ref{fig:setup} right. The profile is fitted by a bi-dimensional-Gaussian function. The fit parameters are used to estimate the area irradiated by the beam and,  therefore, to measure the rate density with the beam flux measurement from the scintillators. The measured beam profile is about few mm$^2$ due to multiple scattering effects on the scintillator mounted upstream.

The angle between the beam and the detectors is chosen to maximize the resolution of the edges in the time distribution of the triple-GEM. An example of the time distribution of the entire hit collection is shown in figure \ref{fig:time_dist} together with a fit with two Fermi-Dirac functions to measure the times at the rising edge and the falling edge. The two parameters are used to estimate the drift velocity of the electrons. The resulting 4.8$\,$cm/$\upmu$s in Ar+10\%iC$_4$H$_{10}$ and 3.7 cm/$\upmu$s in Ar+30\%CO$_2$, are in agreement with the simulations.

\subsection{Space charge effects on gain}

\begin{figure}[tbp]
    \centering
        \begin{tabular}{cc}
        \includegraphics[width=0.4\textwidth]{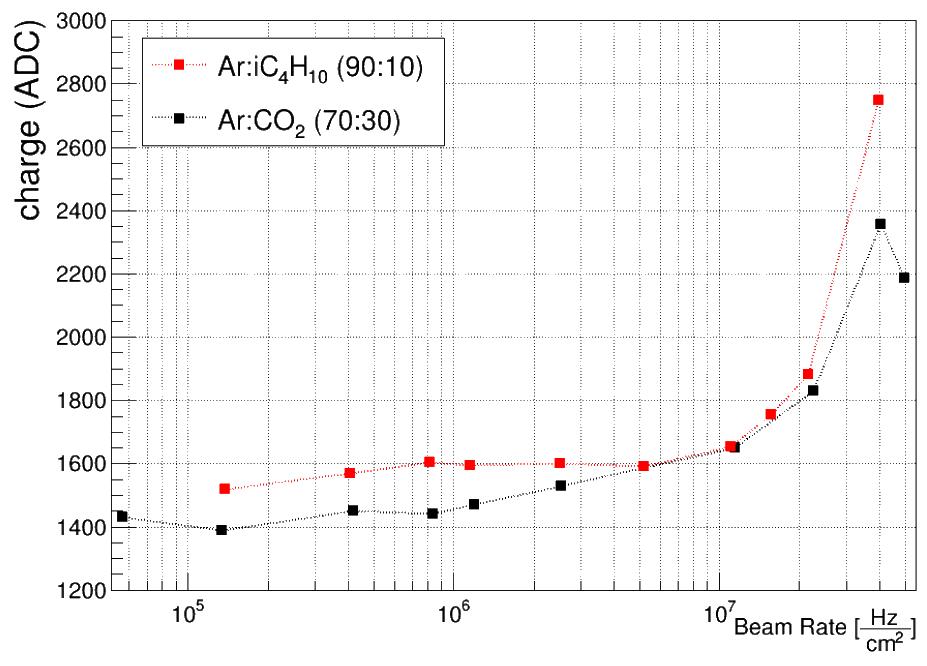} &
        \includegraphics[width=0.47\textwidth]{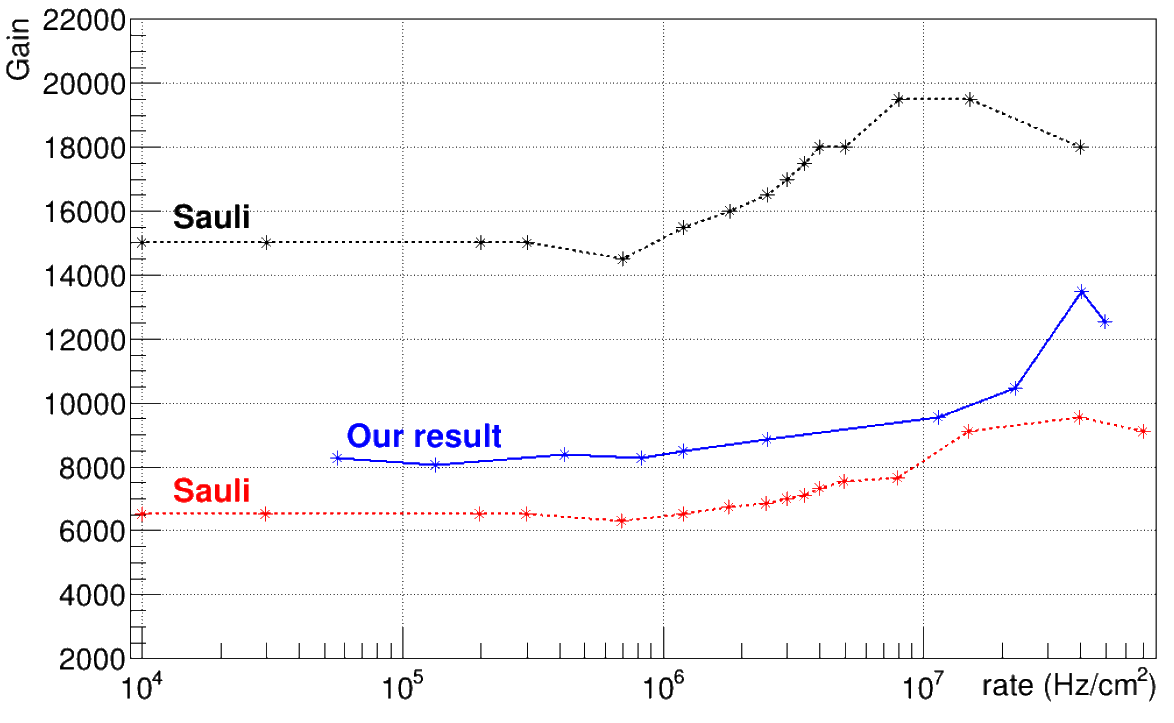}        
        \end{tabular}
        \caption{On the left, mean charge as a function of the beam rate in a triple-GEM with 30$^\circ$ angle between its normal and the beam direction \cite{lavezzi}. On the right, the same points converted in a gain value compared with a similar measurement performed by \cite{charging_up}.}
        \label{fig:charge}
\end{figure}

Triple-GEM technology is well know for its robustness and high rate tolerance. It is know from literature that its gain is stable up to about 10$^8 \,$Hz/cm$^2$, while other technologies such as wire chambers reach values of 10$^5$-10$^6$ Hz/cm$^2$. The GEM rate capability is due to its amplification technique. If the multiplication factor is large, then the electron density inside the hole can affect the electric field. When the number of electrons is higher than the Raether limit \cite{rather} a discharge occurs. 
A measurement of the stability of the detector gain as a function of the beam rate is performed by looking at the cluster charge as function of the beam rate. As it is shown in figure \ref{fig:charge} left, the detector gain is almost flat up to 10$^7$ Hz/cm$^2$, while above this limit space-charge density effects are observed. These results are in agreement with other studies \cite{charging_up} and illustrated in figure \ref{fig:charge}, right where a comparison of results performed with a different setup and configuration are shown. The increase of gain is explained with the charging up of the kapton of the GEM foil: this affects the transparency of the GEM foil and thus its effective gain \cite{thuner}.

\subsection{Space charge effects on $\upmu$TPC}
\begin{figure}[tbp]
    \centering
        \includegraphics[width=0.6\textwidth]{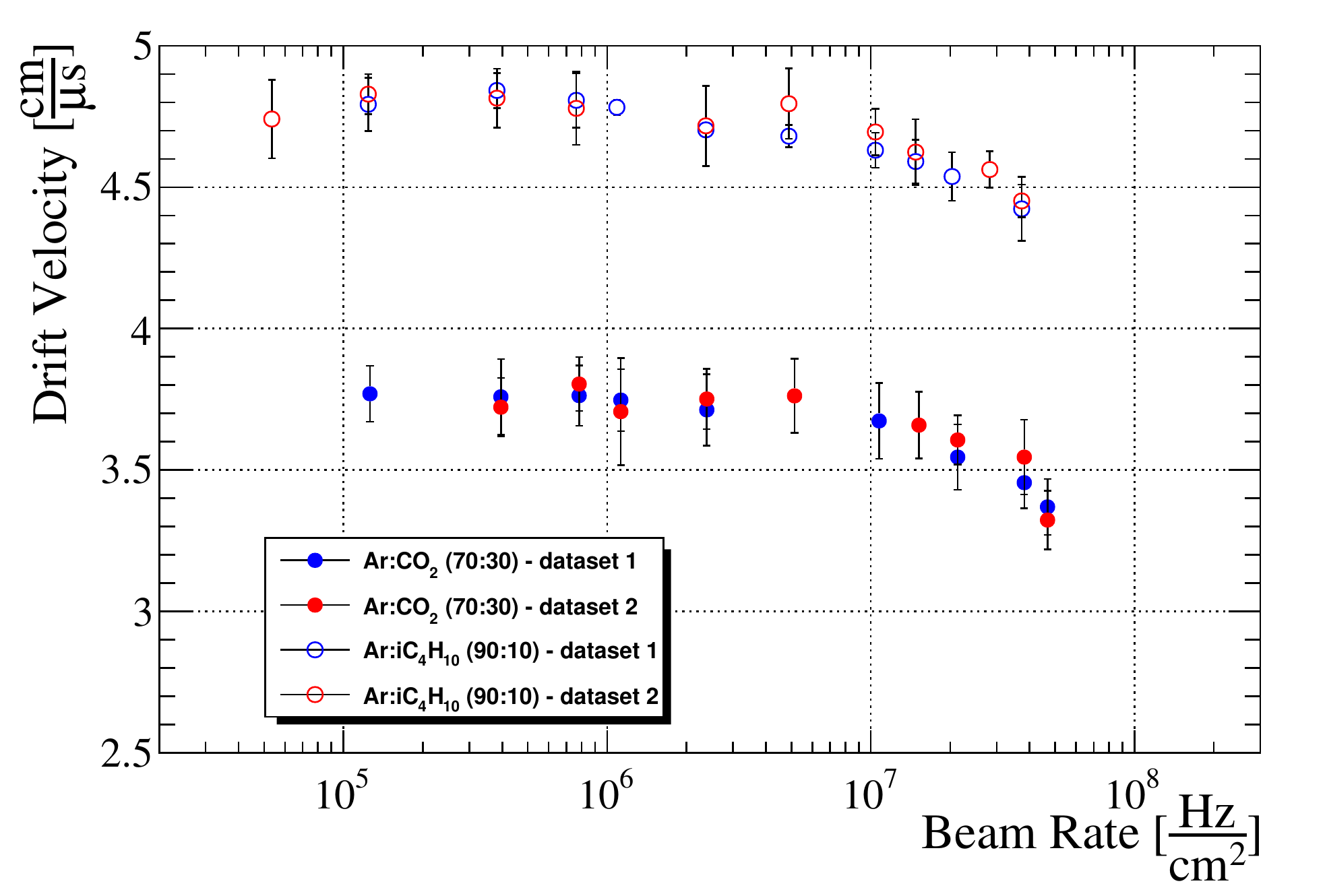} 
        \caption{Measured drift velocity in Ar+10\%iC$_4$H$_{10}$ and Ar+30\%CO$_2$ as a function of the beam rate in a triple-GEM with 30$^{\circ}$ angle between its normal and the beam direction.}
        \label{fig:velocity}
\end{figure}

The space-charge effect does not only affect the effective gain of the GEM, but also the drift properties of the detector. Using the time information of the fired strips, it is possible to evaluate the drift time of the electrons inside the drift gap, as described in section \ref{sec:time}. Similarly to the detector gain, effects on the drift velocity are not observed up to 10$^7 \,$Hz/cm$^2$, but above this value the measured drift velocity drops using either of the mixtures Ar+10\%iC4H10 and Ar+30\%CO2. This study is shown in figure \ref{fig:velocity} with different sets of data. The modifications in the time distribution, therefore, in the drift velocity, are relevant only in the drift gap; this means the observed effect is mainly related to the electrical distortion in the first GEM. A larger effect can take place in the other GEMs since the number of electrons (and ions) increases with the number of amplification stages.
These results put constraints on the GEM performance connected to the beam rate, especially, when the time information is required for the reconstruction. Indeed, since the effective drift velocity depends on the beam rate, the $\upmu$TPC method does not provide a stable spatial resolution. For rates higher than the critical value, the detector cannot profit of this reconstruction method. 

\vspace{0.5cm}
\section{Conclusions}
A triple-GEM detector was tested with an electron flux above 5$\cdot$10$^7 \,$Hz/cm$^2$. An analysis of the performance was performed and a strong degradation of the collected charge and of the drift property of the electrons has been observed for a flux greater than 2$\cdot$10$^7 \,$Hz/cm$^2$. The results concerning the charge are in agreement with previous measurements. The source of the space-charge density effect is related to the charging up of the kapton foil within the two copper faces in a GEM. The dependence of the drift properties on the flux is reported here for the first time and it sets an operational limit for a triple-GEM that uses the $\upmu$TPC algorithm.

\vspace{0.5cm}

\acknowledgments
This work is supported by the Italian Institute of Nuclear Physics (INFN). \\
The research leading to these results has been performed within the FEST Project, funded by the European Commission in the call RISE-MSCA-H2020-2020.


\end{document}